\documentclass[conference]{IEEEtran}

\usepackage{graphicx}
\usepackage{footnote}
\usepackage{caption}
\usepackage{mathtools}
\usepackage{xcolor}
\usepackage{algorithm,algcompatible,amsmath}

\usepackage{array}
\usepackage{tabularx,ragged2e,booktabs}
\usepackage{subcaption}

\begin{document}

\title{Efficient Discovery of Variable-length Time Series Motifs with Large Length Range in Million Scale Time Series}

\author{
    \IEEEauthorblockN{Yifeng Gao\IEEEauthorrefmark{1}, Jessica Lin\IEEEauthorrefmark{1}}
    
    \IEEEauthorblockA{\IEEEauthorrefmark{1}Department of Computer Science, 
     George Mason University, Virginia, USA
     \\
     \{ygao12, jessica\}@gmu.edu}

}

\maketitle

\begin{abstract}
Detecting repeated variable-length patterns, also called variable-length motifs, has received a great amount of attention in recent years. Current state-of-the-art algorithm utilizes fixed-length motif discovery algorithm as a subroutine to enumerate variable-length motifs. As a result, it may take hours or days to execute when enumeration range is large. In this work, we introduce an approximate algorithm called \textbf{H}ierarch\textbf{I}cal based \textbf{M}otif \textbf{E}numeration (HIME) to detect variable-length motifs with a large enumeration range in million-scale time series. We show in the experiments that the scalability of the proposed algorithm is significantly better than that of the state-of-the-art algorithm. Moreover, the motif length range detected by HIME is considerably larger than previous sequence-matching based approximate variable-length motif discovery approach. We demonstrate that HIME can efficiently detect meaningful variable-length motifs in long, real world time series.
\end{abstract}

\IEEEpeerreviewmaketitle

\section{Introduction}

The task of finding repetitive similar patterns in time series data, known as motif discovery, has received a great amount of attention in the past decade. It has been used as an important subroutine in many time series data mining tasks and applications such as association rule mining \cite{shokoohi2015discovery}, data visualization \cite{senin2014grammarviz}, classification \cite{wangrpm}, clustering \cite{buza2010motif}, anomaly detection \cite{senin2014grammarviz}\cite{keogh2005hot}, and activity recognition \cite{minnen2006discovering}. 

Most existing motif discovery algorithms \cite{chiu2003probabilistic}\cite{mueen2009exact}\cite{lonardi2002finding}\cite{lin2007experiencing} consider the length of motif as an important domain knowledge that needs to be determined by user \cite{zhumatrix}. We argue that finding motifs of different, previously unknown lengths---a topic that has been largely glossed over in the literature---is a significant obstacle that prevents motifs from achieving their full potential as a useful time series data mining primitive. 

To demonstrate the limitations of fixed-length motif discovery, we show an example on a subset of a dishwasher electric power demand time series (Fig. 1.top). A basic wash cycle takes approximately half an hour (approximately 350 sample points). By setting motif length equal to one basic cycle, we can find a frequently repeating pattern, representing a wash cycle, as shown in Fig. 1 (Motif A). However, after running our algorithm, we also found one frequent motif (Motif B) of length 1534 (approximately 2.5 hours), and a rare motif (Motif C) of length 2791 (approximately 4.6 hours) that only happened a few times in the time series. The discovery of these two long motifs is surprising given their respective lengths compared to that of the basic wash cycle. Since the lengths of these three motifs are significantly different, fixed-length motif discovery algorithms cannot discover all three motifs in a single run. In contrast, all three motifs are found by our algorithm with just several seconds of execution time. Previous work also note that variable-length motifs are valuable features for many time series data mining tasks \cite{wangrpm}\cite{itr}\cite{mueen2013enumeration}.

\begin{figure}[h]
 \centering
 \vspace{-1mm}
 \includegraphics[width=87mm]{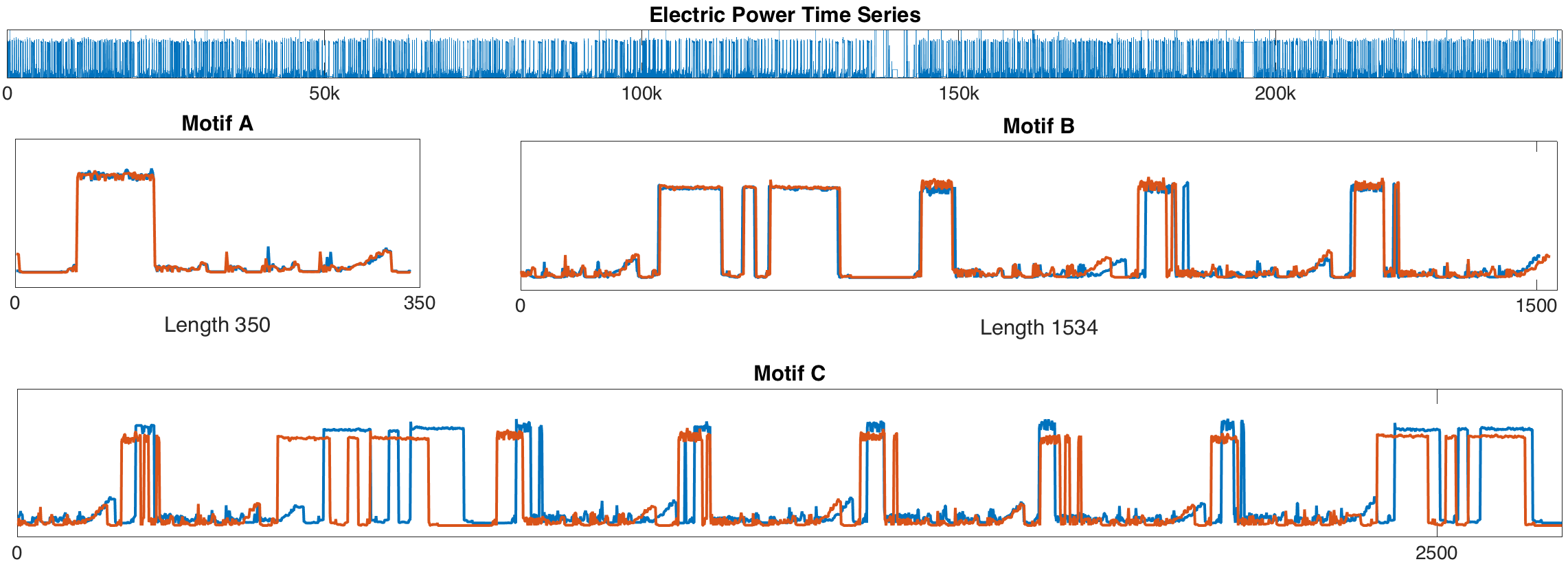}
 \caption{Motifs of different lengths detected by our algorithm in a snippet of dataset from \cite{murray2015energy}.} %Change Later
 \label{fig:1}
 \vspace{-2mm}
\end{figure}

The greatest challenge in detecting variable-length motifs is scalability. We demonstrate this problem by finding motifs of lengths ranging from 300 to 10300 in a ten-million length time series. The brute force algorithm requires $5\times10^{18}$ distance calls to solve this problem. In this case, the computational complexity is similar to that of finding \textit{fixed}-length motif in a 1-billion length time series. In other words, the computational complexity of variable-length motif discovery is 10 times larger than the current largest fixed-length motif discovery problem solved in \cite{zhumatrix} (which takes 12.13 days with GPU speed-up). As a result, even though the state-of-the-art algorithm can achieve 95\% pruning rate, enumerating motifs from lengths 64 to 1024 in an EEG time series of length 160,000 still takes 16 hours  to complete \cite{mueen2013enumeration}.

Million-scale time series exists in many domains. For example, a 2-year power demand dataset used in \cite{murray2015energy} is about 7 million in length. The EOG time series recorded in \cite{goldberger2000physiobank} is about 8 million in length. In both datasets, the state-of-the-art algorithms are too costly to be applied. 
A series of approximate variable-length time series motif discovery algorithms based on grammar induction have been introduced \cite{li2012visualizing}\cite{senin2014grammarviz} following a three-step process. In the first step, the time series is converted to discrete string sequence via a sliding window. In the second step, grammar induction (e.g. via Sequitur \cite{nevill1997identifying}) is applied on the discretized time series to quickly detect repeated string sequences. The third step maps the repeated strings back to the time series subsequences that the strings represent. While the scalability for the grammar-based approach is considerably better than that of the state-of-the-art algorithms, the length range of motifs detected could still be limited. This is because long motifs are often discretized into unnecessarily long and non-identical string sequences due to the typically high amount of noise in the data, and since grammar induction relies on identical sequence matching, existing grammar-based algorithms have difficulty finding long motifs. 

In this paper, we introduce a greedy algorithm named \textbf{H}ierarch\textbf{I}cal based \textbf{M}otif \textbf{E}numeration (HIME) for detecting variable-length motifs. Given a minimum motif length $l$, HIME automatically finds motifs that are (much) larger than $l$ based on a symbol table which records discretized representation of variable-length subsequences. While the algorithm is not an exact algorithm, in the experiments, it can find motifs of lengths from 300 to 3000 in a time series of length one million, with execution time of about 1 minute. This is considerably faster than the state-of-the-art algorithms. Such good scalability makes searching for variable-length motifs in million scale time series a feasible task. Moreover, compared with sequence matching based motif discovery approaches such as \cite{li2012visualizing}\cite{itr}, HIME considerably increases the motif enumeration range that can be detected.

To summarize, our work has the following unique features:

\begin{itemize}

    \item It can discover motifs of much larger length range that sequence matching based approaches \cite{li2012visualizing}\cite{itr} fail to find.
    
    \item The scalability is significantly better than state-of-the-art algorithms.

\end{itemize}

The rest of the paper is organized as follows: Section II discusses related work in fixed- and variable-length motif discovery. Section III introduces the problem definition and notations used in this paper. Section IV introduces our proposed HIME algorithm. An adaptive parameter selection approach is described in Section V. The experimental results are shown in Section VI, and we conclude in Section VII.
\vspace{-2mm}
\section{Related Work}
\vspace{-1mm}
While the classic time series motif discovery problem focuses on finding the most frequent patterns \cite{lin2007experiencing}, much of recent research has defined motifs as the most similar \textit{pair} of subsequences (which we refer to as pair-motifs hereinafter) \cite{mueen2010online}\cite{mueen2009exact}\cite{shokoohi2015discovery}\cite{li2015quick}. In addition, some research work has focused on fast  approximate motif discovery \cite{chiu2003probabilistic}\cite{meng2008mining}. The motifs discovered by these algorithms may contain subseqeunces that are not truly similar. These false positive subsequences can be removed during post-processing. Likewise, similar subsequences may be missed.

The authors in \cite{li2015quick} introduce an algorithm named Quick-Motif, which achieves 3 orders of magnitude speed up compared with the traditional state-of-the-art fixed-length motif discovery algorithm \cite{mueen2009exact}. In a recent work, \cite{yeh2016matrix} introduces an anytime algorithm called STAMP which utilizes fast similarity search algorithm \cite{FastestSimilaritySearch} to find exact pair-motif of a given length. In \cite{zhumatrix}, the authors further introduce an algorithm called STOMP which can reduce the time complexity of STAMP by $O(log(N))$. In \cite{chiu2003probabilistic}, the authors utilize random projection to detect fixed-length approximate motifs. The work in \cite{begum2014rare} focuses on detecting fixed-length approximate motifs with limited memory.

Some recent research work has focused on efficiently finding variable-length motifs. In \cite{nunthanid2011discovery}, the authors introduce an algorithm named VLMD to find exact K pair-motifs by calling fixed-length exact motif discovery algorithm (i.e. MK \cite{mueen2009exact}) with all possible lengths within a range. In \cite{mueen2013enumeration}, the authors introduce an algorithm called MOEN which uses a lower-bound to speed up the process of enumerating motif length. Other works such as \cite{mohammad2014exact}\cite{mohammad2014scale} follow the similar idea of \cite{mueen2013enumeration}. However, a common drawback for these algorithms is that they all conduct an incremental enumerating process starting from the smallest length. Since finding fixed-length motifs is already very costly in million-scale time series, such an approach may be impractical for large-scale variable-length motif discovery.

In \cite{li2012visualizing}\cite{senin2014grammarviz}, the authors introduce a framework, Grammar Induction based motif discovery, which uses grammar induction to find approximate motifs of variable lengths via identifying repeated discrete sequences without enumerating all possible lengths. It is considerably faster than existing algorithms. However, since the idea of the algorithm is based on symbolic sequence matching, the length range of motifs detected may be limited for some real world applications (e.g. the approach cannot detect motifs shown in Fig. 1 in a single run).

Other works such as \cite{tang2008discovering}\cite{castro2010multiresolution} also focus on variable-length motif discovery; however, these techniques have some shortcomings that prevent them from being useful in general cases. In \cite{tang2008discovering}, the subsequences are not normalized, which makes it difficult to find patterns that are similar but with different amplitudes. The work proposed in \cite{castro2010multiresolution} also consists of a discretization step, but the subsequences are non-overlapping. As a result, it does not consider every possible pattern candidate, and thus may miss some true patterns \cite{li2012visualizing}\cite{senin2014grammarviz}.
\vspace{-2mm}
\section{Notations \& Problem Definition}
\vspace{-1mm}
We start with fundamental definitions related to time series:

\textbf{Time Series} $T=t_1,\dots,t_N$ is a set of observations ordered by time. 

\textbf{Subsequence} $s_{p,q}$ of a time serie $T$ is a contiguous set of points in time series $T$ starting from position $p$ and ending at $q$ with length $n=q-p+1$. Typically, $n << N$ and $1\leq p \leq N-n+1$.

Subsequences can be extracted from $T$ via a sliding window. 

In many applications, we are interested in finding similar ``shapes.'' Therefore, motif discovery is more meaningful when it is offset- and amplitude-invariant. This can be achieved by normalizing each subsequence prior to the search for motifs.

\textbf{z-normalization} is a procedure that normalizes the mean and standard deviation of the subsequence to zero and one, respectively.

Given a time series $T$ and a distance function $D(.,.)$, in this work, we define a \textbf{Time Series Motif} to be a set of subsequences in $T$ such that the distance between each of these subsequences and a seed subsequence $s_{seed}$ is less than a motif threshold function $R(L)$, where $L$ is the length of the motif detected, $L \geq l$ (the minimum motif length). Each subsequence in a motif is said to be an instance of the motif.

More specifically, given a minimum motif length $l$, the algorithm finds pairs of subsequences with length $L$ ($L$ is varied for different pairs), where $L \geq l$ and each pair satisfies motif definition with threshold $R(L)$. 
Note that as mentioned in \cite{begum2014rare}, once the pair is detected, all instances of motif can be found by using similarity query algorithms \cite{FastestSimilaritySearch}\cite{rakthanmanon2012searching}.

\vspace{-3mm}
\subsection{Problem Definition}

Different from fixed-length motif discovery, determining a general measure (e.g. frequency or similarity) for variable-length motif discovery problem is non-trivial \cite{li2012visualizing}\cite{nunthanid2011discovery}. For example, some low frequency motif may be more interesting than high frequency motif; some long motifs are likely to have a large distance compared with short motifs even after normalizing by length (e.g. Motif C versus Motif A in Fig. 1). Since motif discovery can be used as a subroutine for real world applications, instead of ranking motifs based on some interest measure, the proposed algorithm aims to detect ``seed subsequence" with at least a pair of subsequences that satisfy the motif constraint stated above. The reader may refer to \cite{mueen2013enumeration}\cite{wangrpm}\cite{senin2014grammarviz} for details on various motif evaluation approaches for different goals. Since most state-of-the-art approaches evaluate motifs on similarity, when compared with them in the experimental section, we include the execution time of ranking the motifs based on similarity \cite{mueen2009exact}\cite{mueen2013enumeration}.

\begin{figure}[h]
 \centering
\vspace{-3mm}
 \includegraphics[width=80mm]{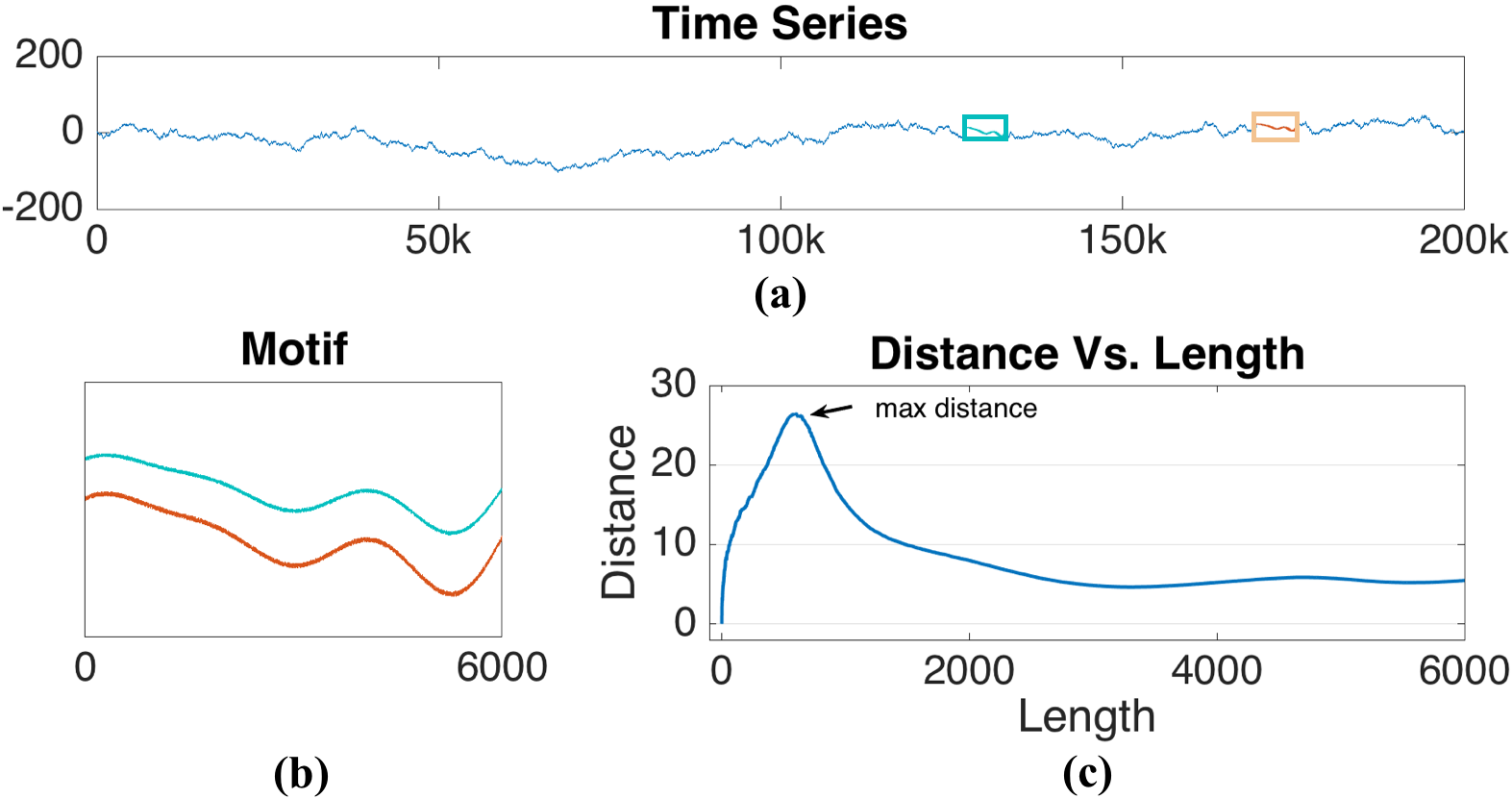}
  \vspace{-2mm}
 \caption{Example of Normalized Distance as Growth of Length}
 \label{fig:2m}
 \vspace{-3mm}
\end{figure}

The difficulty of getting exact solution for variable-length motif can be illustrated by a simple example shown in Fig. \ref{fig:2m}. Suppose two instances of a motif, shown in Fig. \ref{fig:2m}(b), appear in a time series (Fig. \ref{fig:2m}(a)) at the $i^{th}$ and $j^{th}$ positions, respectively. Fig. \ref{fig:2m}(c) shows the growth of normalized Euclidean distances between subsequences $s_{i,i+n-1}$ and $s_{j,j+n-1}$ for all lengths $n$ ranging from 2 to 6000. We can see from Fig. \ref{fig:2m}(c) that the growth of the  distance is nonlinear. That is, the distance between a pair of short subsequences does not necessarily share similar behavior with the distance between long subsequences if the length difference is large. Therefore, the state-of-the-art algorithms need to repeat the fixed-length motif discovery algorithm many times during the enumeration process to find optimal motifs, which significantly increases the time cost. 
\vspace{-2mm}
\section{Proposed Method}
\vspace{-2mm}

In this section, we introduce our proposed method.
\subsection{Discretization}

\begin{figure}[h]
 \centering

 \vspace{-3mm}
 \includegraphics[width=75mm]{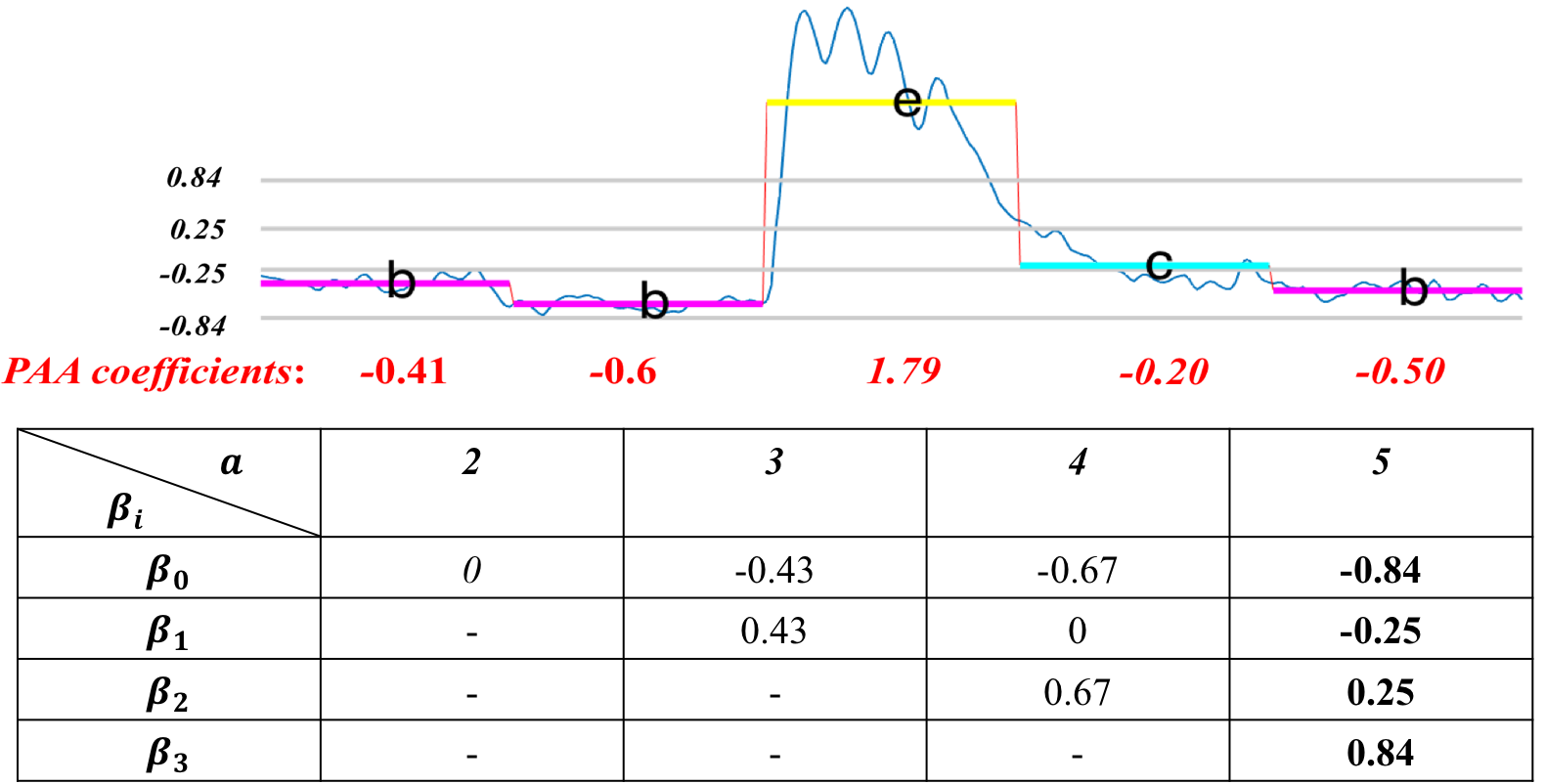}
 \caption{Example of Generating SAX word}
 \label{fig:sax:example}
 \vspace{-5mm}
\end{figure}

Discretization of time series into symbolic representation is often a necessary pre-processing step for efficient motif discovery \cite{lonardi2002finding}\cite{chiu2003probabilistic}\cite{senin2014grammarviz}\cite{begum2014rare}. Since our proposed work also utilizes symbolic representation of subsequences to speed up motif detection process, we first describe Symbolic Aggregate approXimation (SAX) \cite{lin2007experiencing}, a widely used discretization technique for time series data mining. Given a z-normalized time series (a subsequence in our case), SAX first converts it to Piecewise Aggregate Approximation (PAA) representation \cite{lonardi2002finding} of size $w$ (i.e. $w$ segments). Then, the PAA coefficients are mapped to $w$ symbols with alphabet size $a$ according to a breakpoints table \cite{lin2007experiencing}, defined such that the regions are approximately equal-probable under Gaussian distribution. These $w$ symbols form a SAX word. Fig. \ref{fig:sax:example} illustrates the discretization process. The pre-defined breakpoints table \cite{lin2007experiencing} for alphabet size up to $a=5$ is shown in Fig. \ref{fig:sax:example} (bottom).

\subsection{Fast Computation of SAX}
 \vspace{-1mm}

Since SAX words are heavily used in the proposed algorithm, it is very important to reduce the time cost of discretization. We introduce a fast way to compute SAX words from subsequences of different lengths. Two vectors of statistical features, $M_x(x)=\sum_{i=1}^{x}t_i$ and $M_{xx}(x)=\sum_{i=1}^{x}t_i^2$ are first computed based on input time series  $T$. Given a subsequence $S_{p,q}$ of length $n$, its SAX representation can be computed by Algorithm 1. The algorithm uses a fast PAA computation approach \cite{li2015quick}\cite{rakthanmanon2012searching} to compute PAA coefficients with time complexity $O(w)$ which is not dependent on subsequence length (Lines 5-7). In Line 8, the PAA coefficients are converted to SAX word based on the pre-defined breakpoints table, which cost $O(wa)$. The cost of Algorithm 1 is thus $O(w+wa))$. Since $M_x(x),M_{xx}(x)$ can be computed during pre-possessing, the cost of computing a SAX word for arbitrary length subsequence during motif discovery process is $O(w+wa)$. As demonstrated in \cite{lin2007experiencing}, $w$ and $a$ should be very small compared with subsequence length. So the time cost to compute a SAX word is reduced from $O(n+wa)$ to $O(w+wa)$ ($w\ll n$).

\begin{algorithm}[h]
    \caption{Fast SAX Computation (FastSAX)}
  \begin{algorithmic}[1]
    \STATE \textbf{Input}: $M_x$,$M_{xx}$, PAA size $w$, subsequence $S_{p,q}$
    \STATE \textbf{Output}: PAA representation $paa$
    \STATE $Ex=M_x(q)-M_x(p)$, $E_{xx}=M_{xx}(q)-M_{xx}(p)$
    \STATE $n=q-p+1$,  $\mu_x=\frac{E_{x}}{n}$, $\sigma_x=\sqrt{\frac{E_{xx}-E_{x}^2/n}{n-1}}$
    \FOR{every PAA segment}
     \STATE $paa_i=(\frac{M_{x}(paa_{i,end})-M_{x}(paa_{i,start})}{n/w}-\mu_x)/\sigma_x$
     \ENDFOR
     \STATE \textbf{return} ConvertToSAX($paa$)
  \end{algorithmic}
\end{algorithm}

\vspace{-2mm}
\subsection{Lower-bound based Numerosity Reduction}

In practice, neighboring subsequences are similar to each other since they are off by one point. To reduce the cost of detecting long motifs, Numerosity Reduction (NR) is used to avoid unnecessary similarity comparisons. Different from previous work \cite{li2012visualizing}\cite{itr}, where a subsequence is skipped if its SAX representation is identical to the last recorded one, we use PAA distance to remove consecutive similar subsequences. A subsequence is ignored if the lower-bounding PAA distance between neighboring subsequences is less than $2R(l)$. Since we want a tight lower bound, we use a large PAA size. In this work, we use PAA size of 32 for numerosity reduction. 

\vspace{-1mm}
\subsection{Induction Graph}
\vspace{-3mm}
\begin{figure}[h]
 \centering
 %\vspace{-1em}
 \includegraphics[width=80mm]{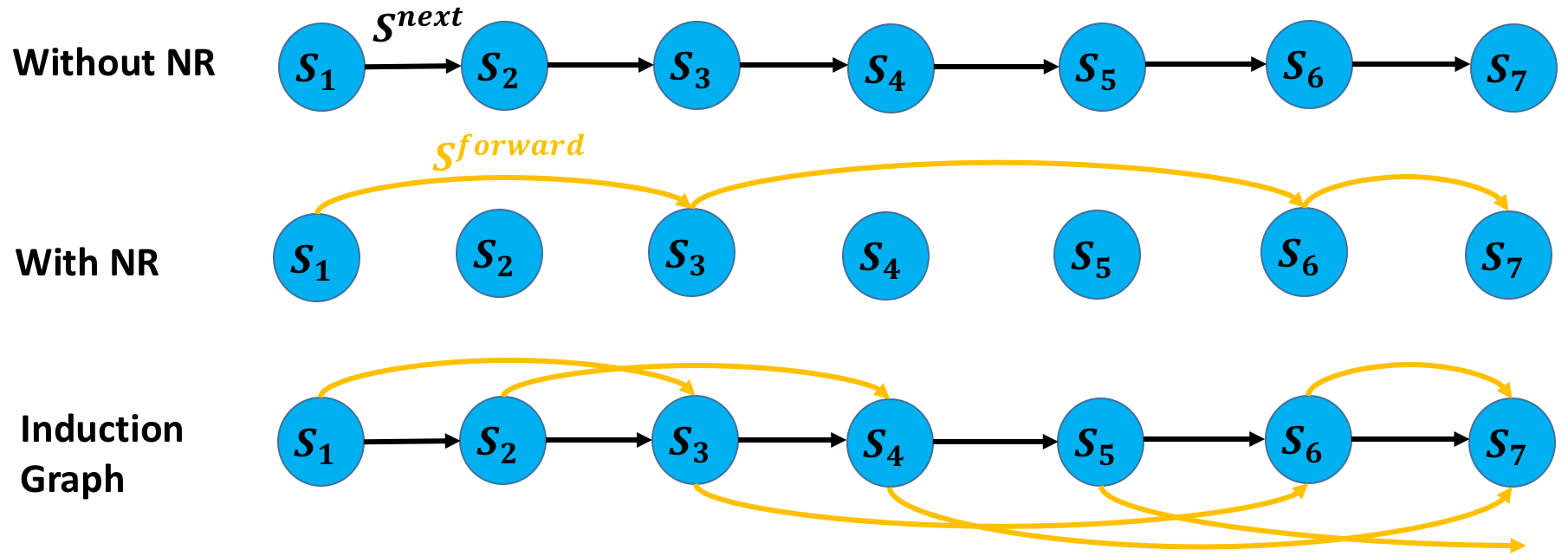}
 \caption{Illustration of Induction Graph}
 \label{fig:indgraph}
 \vspace{-3mm}
\end{figure}

 The number of SAX words that need to be tested to form long motifs can be unnecessarily long if numerosity reduction is not employed. See Fig. \ref{fig:indgraph} (top). The time series is converted to a sequence of SAX words: $S_1$, $S_2$, ..., $S_7$. The $i^{th}$ node in the graph represents the subsequence $s_{i,i+l-1}$ in time series. Without numerosity reduction, every single subsequence/SAX word is kept. However, numerosity reduction may cause some patterns to be missed since some subsequences are skipped. In the example shown in Fig. \ref{fig:indgraph} (middle), $S_2$, $S_4$ and $S_5$ are skipped due to numerosity reduction, and some patterns involving the skipped subsequences may be missed. To mitigate the drawbacks of both scenarios, we introduce the Induction Graph (Fig. \ref{fig:indgraph} (bottom)), a graph structure that helps determine the order of scanning and enumeration of motif candidates during motif discovery. Each node contains 2 edges (next edge and forward edge). Two nodes connected by these two edges are denoted as $S_i^{next}$ and $S_i^{forward}$  (In Fig. \ref{fig:indgraph} (bottom), these are the black and yellow arrows, respectively). $S_i^{next}$ is the node representing the subsequence $s_{i+1,i+l}$ and $S_i^{forward}$ is the next non-similar subsequence determined by numerosity reduction (see previous section). The Induction Graph can be stored by only recording the nodes connected by $S_i^{forward}$. For example, in Fig. \ref{fig:indgraph}, all edges and nodes in the Induction Graph can be reconstructed if we record 4 nodes: $\textbf{S}_{NR}=S_1,S_3,S_6,S_7$. In the rest of the paper, the nodes connected by reverse direction of next edge and forward edge are denoted as $S_i^{prev}$ and $S_i^{backward}$ respectively.

\subsection{Hierarchical based Motif Enumeration (HIME)}

Hierarchical based Motif Enumeration (HIME) is described in Algorithm 2. Intuitively, the algorithm conducts a left to right passing through all nodes via the next edge in Induction Graph $G$. For each node, the algorithm recursively executes two major functions: ``RecursiveEnumeration'' (Lines 16-17) and ``RemoveCoveredMotif'' (Line 13). In ``RecursiveEnumeration'' step, SAX words are formed to represent variable-length subsequences and to detect repeating subsequences. ``RemoveCoveredMotif'' removes short motifs found that are completely covered by longer motifs to maintain a small size of motif set at a low cost. The motifs detected by the algorithm are stored in $MotifSet$ (Line 12). Finally, post-processing is applied to remove trivial and false-positive motifs candidates (Line 20). Note that HIME can also utilize numerosity reduction by only examining the nodes recorded in $\textbf{S}_{NR}$ of $G$. 

\begin{algorithm}[h]
    \caption{Hierarchical based Motif Enumeration (HIME)}
  \begin{algorithmic}[1]
    \STATE \textbf{Input}: Induction Graph $G$
    \STATE \textbf{Parameter}: PAA size $w$, Alphabeta Size $a$
    \STATE \textbf{Output}: Motif Set $MotifSet$
    \STATE VLSAXTable[SAX word, Length][Location]=\{\}
    \FOR{each node $S_i$ from left to right}
    
       \COMMENT{Compute SAX word $word$ for long subsequence}
    
       \STATE $SS$=Merge($S_i$,$S_i^{forward}$); $word$=FastSAX($SS$,$w$,$a$);
       
       \COMMENT{Check if the same SAX word representing some subsequence with similar length is recorded}
       \IF{!VLSAXTable.exist($word$,$SS$.Length)}
         \STATE VLSAXTable.put($word$,$SS$.Length,$SS$.Location);
       \ELSE
         
         \COMMENT{Retrieve subsequence with matching SAX word}
         \STATE $SS2$=VLSAXTable.getSimLengthSeq($word$);
         
         \COMMENT{Update Induction Graph $G$}
         \STATE InsertMotifNode($G$,$SS2$,$SS$);
         \STATE UpdateMotifSet(MotifSet,$word$);
         
         \COMMENT{Greedy Removing Covered Motifs}
         \STATE RemoveCoveredMotif($SS$,$S_i$);
      
         \COMMENT{Enumerate Motifs based on $SS$ and $SS2$}
          \STATE $SS3$=Merge($SS$,$SS^{forward}$);
          \STATE $SS4$=Merge($SS2$,$SS2^{forward}$);
          
      \STATE RecursiveEnumeration($SS3$);
      \STATE RecursiveEnumeration($SS4$);
      \ENDIF
    \ENDFOR
    \STATE \textbf{return} RemoveTrivialAndFalsePostive(MotifSet);
  \end{algorithmic}
\end{algorithm}

\subsubsection{SAX based Recursive Enumeration}

In this step, Algorithm 2 recursively executes Lines 6-18 to detect variable-length motifs. First, HIME computes a SAX word $word$ and generates a new node $SS$ to represent the long subsequence obtained by merging two short subsequences $S_i$ and $S_i^{forward}$ (Line 6). The new SAX word $word$, along with the total length of the merged subsequence and the location of the subsequence are inserted into a SAX word table, VLSAXTable, if the same SAX word representing some subsequence(s) of similar length does not already exist in the table (Lines 7-8). If $word$ already exists in VLSAXTable for some subsequence(s) of similar length, we have found a motif match. The algorithm gets $SS2$ based on the location of the matching subsequence (Line 10) and inserts $SS$ into the graph (Line 11). $SS$ and $SS2$ are now instances of a motif. The \textit{next} edge and \textit{forward} edge of $SS$ are updated based on $S_i^{forward}$'s \textit{next} edge and \textit{forward} edge respectively. The reverse links are connected to $S_i^{backward}$ and $S_i^{prev}$ respectively. The insertion of the new nodes allows us to re-use the detected motifs to reduce the cost of enumerating long motifs. The algorithm then recursively tests two new generated long subsequences represented by $SS3$ and $SS4$ (generated by further merging $SS$ with $SS^{forward}$, and $SS2$ with $SS2^{forward}$, respectively) until Line 7 is satisfied (Lines 16-17). Such greedy recursive strategy can efficiently generate a structure of low hierarchy for matching long repeating patterns \cite{nevill1997identifying}. 

\begin{figure}[h]
 \centering
 \vspace{-2mm}
 \includegraphics[width=75mm]{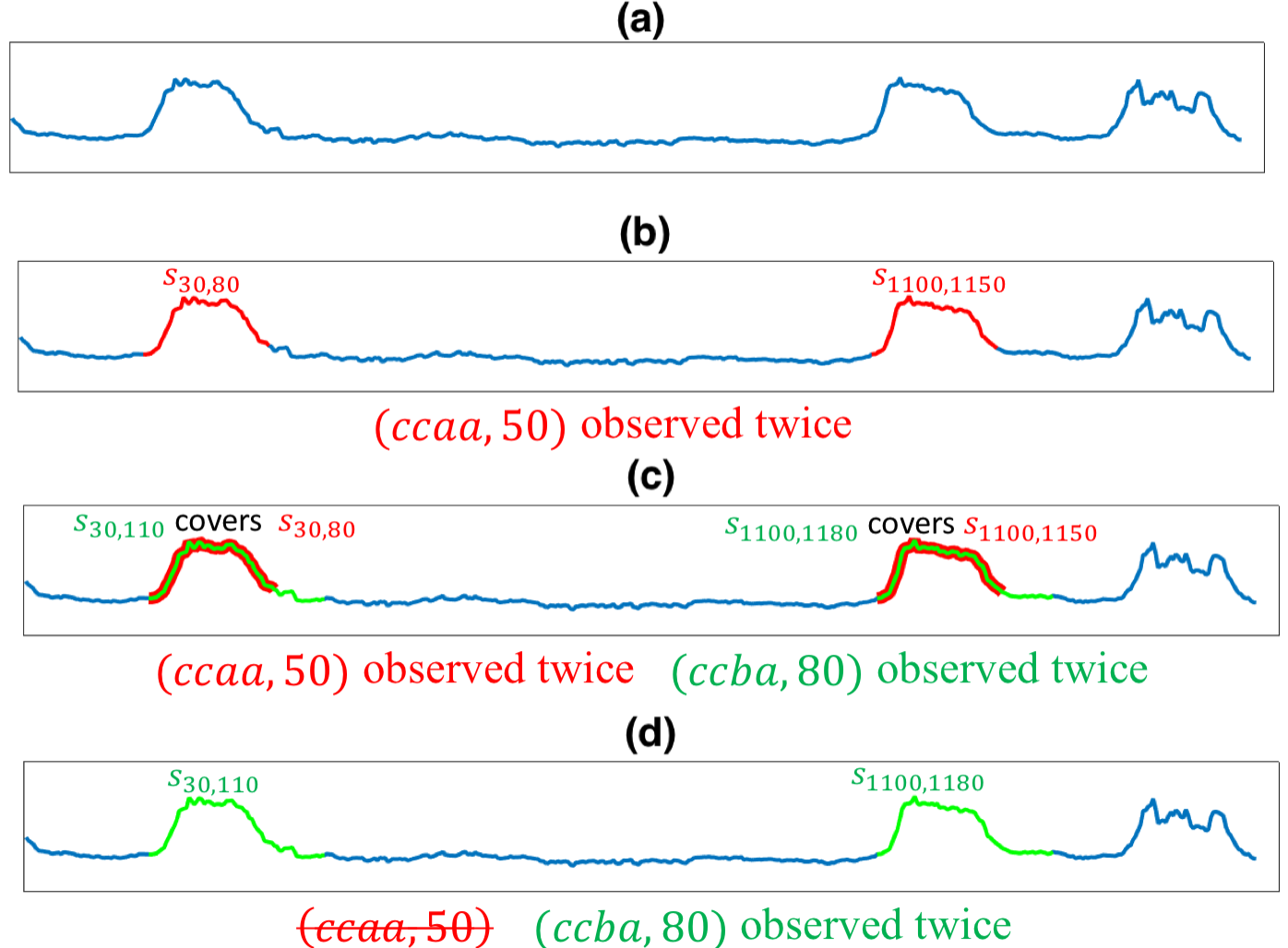}
 \caption{Example showing the greedy removal of covered motif}
 \label{fig:cov}
  \vspace{-2mm}
\end{figure}

\subsubsection{Removing Covered Motifs}

Some short motifs may completely overlap with long motifs. As demonstrated in \cite{nunthanid2011discovery}\cite{mueen2013enumeration}, these short motifs (``covered motifs'') are redundant. So during the motif enumeration process, we conduct a fast process (Line 13) to remove the entries for covered motifs from VLSAXTable and MotifSet. An example is shown in Fig. \ref{fig:cov} to illustrate the process. The $(word,L)$ pair (e.g. $(ccaa,50)$) in the figure denotes the SAX word $word$ that represents a subsequence of length $L$. The subsequences having the same $(word,L)$ pair are instances of the same candidate motif, and are labeled in the same color in the figure. In the example, the algorithm first finds the short motif (Fig. \ref{fig:cov}(b)) based on $(ccaa,50)$ ($s_{30,80}$ matches $s_{1100,1150}$). The algorithm then expands $s_{1100,1150}$ using the recursive enumeration strategy mentioned above and forms a longer subsequence represented by $(ccba,80)$. A match $s_{30,110}$ is also found (labeled in green in Fig. \ref{fig:cov}(c)). Since $(ccaa,80)$ is generated from $(ccaa,50)$ and they both represent the same subsequence (except one is a longer version), all instances of the motif represented by $(ccaa,50)$ may overlap with all instances of long motif represented by $(ccaa,80)$. Therefore, the algorithm removes subsequences associated with $(ccaa,50)$ from VLSAXTable and the corresponding candidate motif from motif set (Fig. \ref{fig:cov}(d)). While ``RemoveCoveredMotif'' cannot remove all covered motif, it is used to minimize the time and space cost during motif discovery. All covered motifs can be removed using the algorithm described in \cite{mueen2013enumeration} after getting the motif set.

\subsection{Compared with Sequence-Matching Based Approximate Motif Discovery Approaches}

Compared with SAX-based sequence matching approximate motif discovery approaches (e.g.  \cite{li2012visualizing}\cite{itr}), HIME has two unique advantages. 

\begin{figure}[h]
 \centering
 \vspace{-3mm}
 \includegraphics[width=75mm]{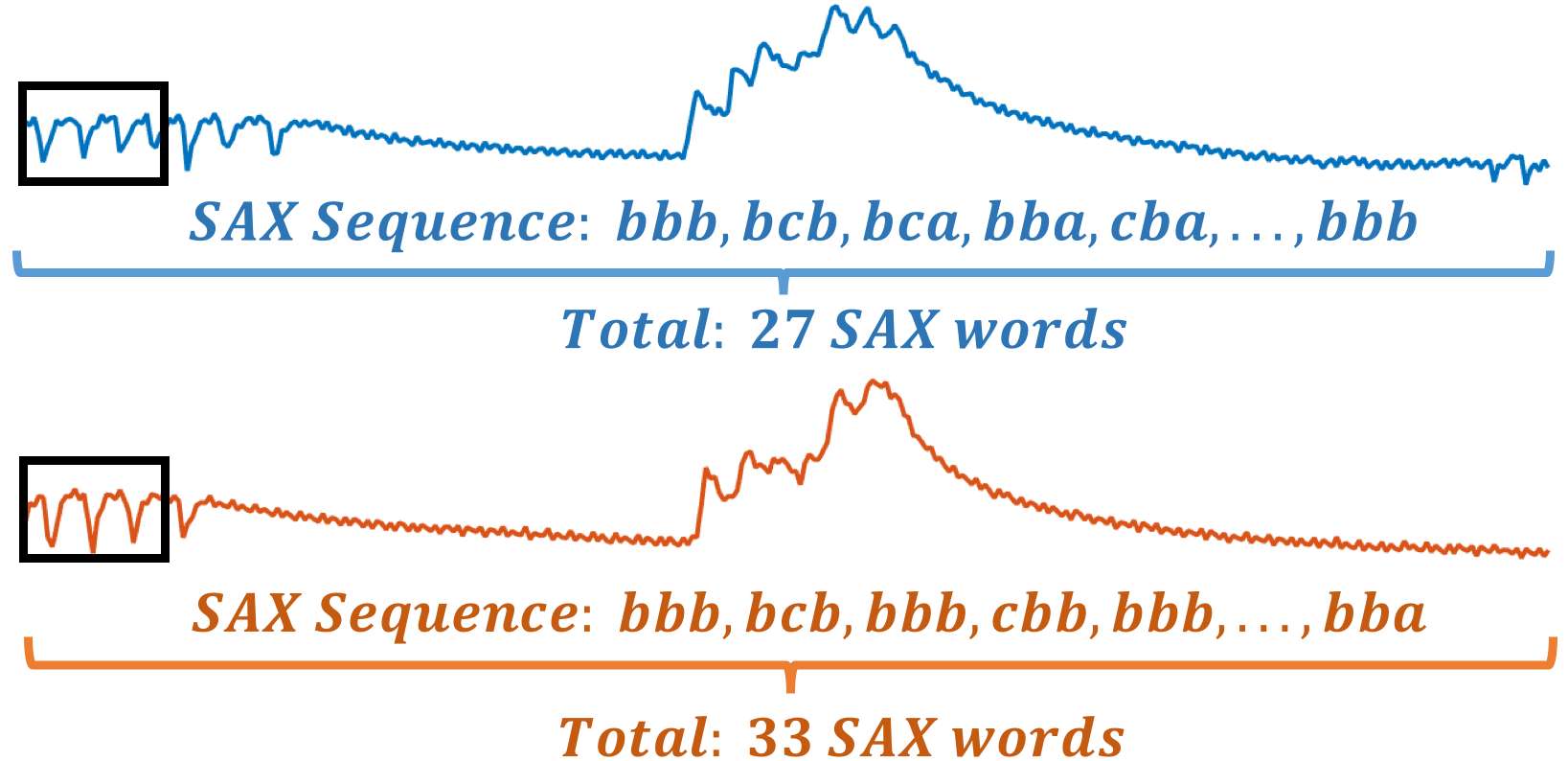}
 \caption{An example showing the difficulty for Grammar Induction approach to find long motifs}
 \label{fig:l}
  \vspace{-4mm}
\end{figure}

First, these existing approaches can detect a long motif only if the SAX sequences representing the motif instances are identical. To illustrate the difficulty of generating the same sequence for long motif, an example is shown in Fig. \ref{fig:l}. The two subsequences of length 500 in Fig. \ref{fig:l} are instances of a motif discovered by our algorithm in EPG time series \cite{mueen2013enumeration}. If we set the minimum length to be 50, these two subsequences form two SAX sequences of lengths 27 and 33, respectively, with numerosity reduction. Clearly, the two subsequences are converted to overwhelmingly long word sequences that are not similar to one another except the first two words. Therefore, sequence-matching based approaches cannot detect them. In contrast, HIME would form a single SAX word for each subsequence via the recursive enumeration process, and find motifs using only the SAX word representing the subsequence. In this example, since the two subsequences indeed have the same SAX representation, HIME has the ability to discover them.

Second, the mean and the variance of a subsequence can affect the shape dramatically \cite{mueen2013enumeration}. However, the generation of SAX words in the word sequence only depends on the mean and the variance of small subsequence. Therefore, when the length of word sequence is long, the mean and the variance of short subsequences may significantly differ from the overall mean and variance of the whole subsequence. So even if the subsequences are similar with each other, their respective word sequences may be dissimilar. HIME avoids this problem by fast re-computing SAX word via Algorithm 1, which re-normalizes the subsequence every time.

In the experiments, we demonstrate that HIME significantly increases the enumeration range compared to sequence-matching based approaches.

\section{Parameter Selection}

\begin{figure}[h]
 \centering
 \vspace{-2mm}
 \includegraphics[width=70mm]{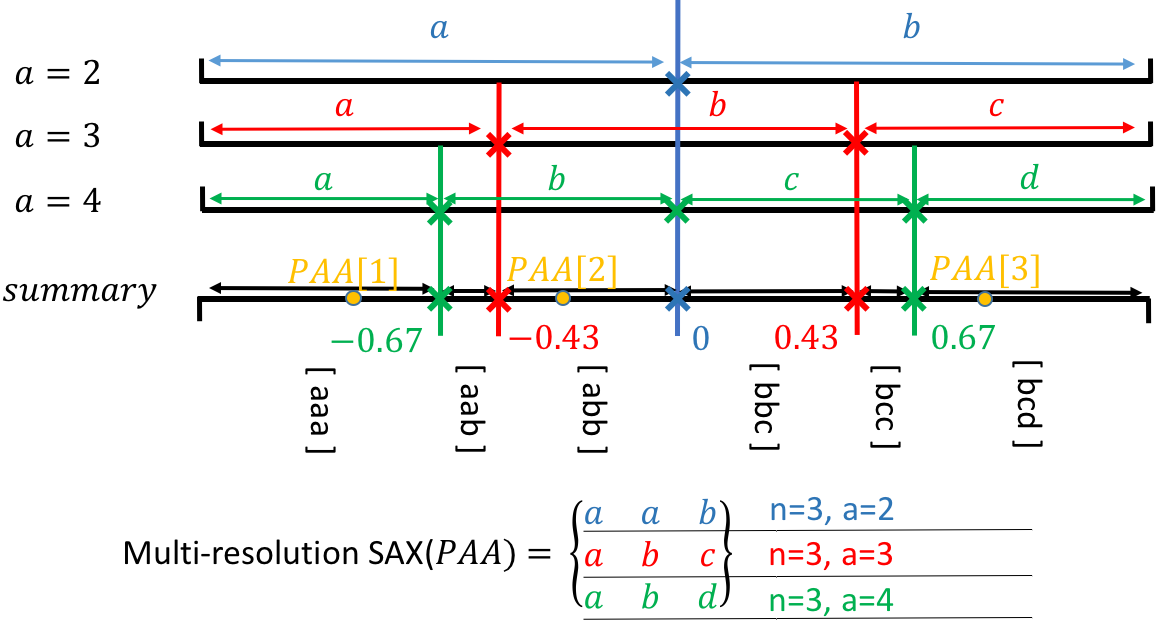}
 \caption{Fast Computing Multi-resolution SAX}
 \label{fig:saxm}
 \vspace{-8 mm}
\end{figure}

Since the performance of the algorithm depends on the SAX parameters $w$ and $a$, in order to alleviate the burden of parameter selection, we introduce an approach to adaptively choose parameter $a$. 

We first introduce a fast approach to compute multi-resolution SAX with similar cost as fixed-resolution SAX. Given a maximum alphabet size $a_{max}$, our approach first gathers all SAX breakpoints for alphabet sizes from 2 up to $a_{max}$. For each interval between any two breakpoints, a symbol sequence containing corresponding symbols up to $a_{max}$ resolution is recorded. An example with $a_{max}=4$ is shown in Fig. \ref{fig:saxm}. The breakpoints are denoted by ``x," and the symbol sequence for each intervals is shown in $[.]$. By using binary search to determine which interval the PAA coefficient belongs to, we can find its SAX representations in all resolutions from $2$ to $a_{max}$ at a low cost. Since the number of intervals is bound by $O({a_{max}}^2)$, the time complexity of binary search is bound by $O(2log(a_{max}))$. Given the PAA values denoted in yellow as an example, the algorithm computes 3 different resolutions of SAX words $aab,abc,abd$ by simply concatenating three symbol sequences $aaa,abb,bcd$ (shown in $\{.\}$). Determining each symbol sequence only takes 3 binary searches. As mentioned in \cite{senin2014grammarviz}, for motif discovery , $a$ generally is chosen under 20, which only contains 128 distinct break points. So the complexity of computing a single multi-resolution SAX with PAA size $w$ word with $a$ up to 20 only cost $O(1+7w)$. 

The Adaptive Parameter Selection Algorithm is outlined in Algorithm 3. The algorithm random-samples a pair of subsequence ($S_1$,$S_2$) of length $l$ until alphabet size $a$ is determined. For each pair sampled, a suitable $a$ that makes the tightness of lower-bound $tightness_{LB}$ (computed by $SAXMINDIST(S_1,S_2)/EuclideanDist(S_1,S_2)$) of SAX word close to 0.5 is recorded. The threshold (0.5) provides a way to balance the tightness of the representation and the number of distinct SAX words. Since $tightness_{LB}$ monotonically increases as $a$ increases \cite{lin2007experiencing}, using binary search (Lines 4-6), we can find the proper $a$ without examining all resolutions. Intuitively, $tightness_{LB}$ controls the approximate performance of SAX words and can be used to select a suitable resolution. The average alphabet size $\bar{a}$ is selected to be the parameter for HIME algorithm after the parameter selection algorithm converges (e.g. when the change of $\bar{a}$ is less than 0.01). By pre-computing SAX words of all resolutions for ($S_1$,$S_2$), BinarySearchResolution in Algorithm 3 only costs $O(5)$ to search suitable $a$ from 2 to $a_{max}=20$. During the experiment, $\bar{a}$ often converges after one thousand samplings, which only takes less than one second.

\begin{algorithm}[h]
    \caption{Determining Alphabet Size for HIME}
  \begin{algorithmic}[1]
    \STATE \textbf{Input}: Time Series $T$
    \STATE \textbf{Output}: Alphabet size $\bar{a}$
    \WHILE{$\bar{a}$ not converged}
      \STATE Random Sampling $S_1$,$S_2$ from time series $T$
      \STATE $a_i$=BinarySearchResolution($S_1$,$S_2$);
      \STATE $\bar{a}$=Average($a$);
     \ENDWHILE
     \STATE \textbf{return} $\bar{a}$
  \end{algorithmic}
\end{algorithm}

\section{Evaluation Experiments}

We perform a series of experiments to evaluate the performance of HIME. All the experiments are conducted on a 16 GB RAM laptop with quad core processor of 2.5 GHz. The executable software and datasets used in the experiments can be found in \textit{http://bit.ly/2rvBETV}. Note that the goal of the experiments is to demonstrate the contributions and potential applications for efficiently finding motifs with large length differences that existing algorithms may have difficulty to find. 

We first test 2 different state-of-the-art enumeration approaches \cite{nunthanid2011discovery}\cite{mueen2013enumeration} in the task of detecting motifs of length from 300 to 2300 in three real-world time series of length 160,000 to demonstrate the scalability problem. The first approach iteratively calls fixed-length motif discovery algorithm (e.g. MK \cite{mueen2009exact}) to find the most similar subsequences in different lengths \cite{nunthanid2011discovery}. We choose the current fastest algorithm, Quick-Motif \cite{li2015quick}, to maximize its scalability (denoted as ItrQuick-Moitf). The second approach \cite{mueen2013enumeration} (MOEN) applies an all pair-wise similarity search algorithm and a lower-bound to prune some lengths that do not need to be tested. We choose the fastest algorithms, STOMP \cite{zhumatrix} and STAMP \cite{yeh2016matrix}, to be used with the MOEN framework. Since STAMP is slower than STOMP in all test cases, we only show the execution time for motif enumeration based on STOMP (denoted MOEN-STOMP). All execution times of STAMP and STOMP are based on the C code provided by the authors. Table I shows the execution time of fixed-length motif discovery algorithm, MOEN (using code provided by the author) and the estimated execution time if Quick-Motif and STOMP are applied in both enumeration approaches respectively (calculated based on the number of times fixed-length algorithm is called in the framework). From the results, ItrQuick-Motif is very costly compared to the MOEN framework. Although MOEN reaches $90\%$, $95\%$ and $98\%$ pruning rates in 3 datasets respectively, even using the latest all-pair similarity comparison algorithm, MOEN-STOMP still takes hours to detect the motifs. In contrast, HIME only take few seconds including post-processing. So state-of-the-art algorithms are hard to be applied in the remaining experiments since the smallest time series used has length of 1 million. Therefore, instead of directly comparing scalability performance, we compare our performance with STOMP $+$ $99\%$ pruning rate. As shown in the table, 99\% pruning is hard to reach even in periodic ECG time series. 

Similar to ItrQuick-Motif, directly using fixed-length \textit{approximate} motif discovery \cite{begum2014rare}\cite{chiu2003probabilistic}\cite{liu2015efficient} is also costly. For example, even if approximate motif discovery algorithm only takes one second to find fixed-length motifs, finding motifs in a length range of [300, 2300] still requires approximately 33 min. In long time series, just the discretization step alone in \cite{chiu2003probabilistic}\cite{liu2015efficient} may take more than one second. Similar problem also exists in fixed-length anytime version of the state-of-the-art motif discovery algorithm, STAMP \cite{yeh2016matrix}. STAMP can only examine fewer than 50 out of 16 million subsequences in one second.

\begin{table}[h]
 \vspace{-1mm}
\caption{Time required to detect motifs of lengths from 300 to 2300 in time series of length 160,000. Some of the times are estimates only.}
\centering
\begin{tabular}{ |c|c|c|c|}
  \hline
   Algorithm$/$DataSet & Electric Power & EEG & ECG \\
   \hline\hline
  STAMP (fixed) & 12.15 min & 12 min & 13.7 min \\
  STOMP (fixed) & 4.15 min & 5.19 min & 4.05 min \\
  Quick-Motif (fixed) & 8.2 min & 7.8 min & 30 sec. \\
  \hline
  MOEN & 3.1 days  & 1.7 days  & 16 hr\\
  ItrQuick-Motif (est.) & \textcolor{gray}{11.3 days} & \textcolor{gray}{10.08 days} & \textcolor{gray}{16.6 hr}\\
  MOEN-STOMP (est.) & \textcolor{gray}{13.8 hr} & \textcolor{gray}{8.65 hr} & \textcolor{gray}{2.7 hr} \\
  \hline
  HIME+Post-Processing & \textbf{23 sec} & \textbf{21 sec} & \textbf{47 sec} \\
  \hline
\end{tabular}
\label{tab:memory}
 \vspace{-3mm}
\end{table}

In all experiments unless noted, the alphabet size is set using Algorithm 3. The PAA size $w$ is set to 6. The minimum length $l$ is 300 and motif threshold is $R(L)=0.02L$. The relation between execution time, enumeration range and parameter setting is discussed in Sec VI.C. 

\vspace{-2mm}
\subsection{Detecting Planted Motifs in Random Walk Time Series}
\vspace{-1mm}
We first test HIME in a planted motif experiment to demonstrate its ability to detect motifs with high accuracy. We choose a grammar (Sequitur) based motif discovery algorithm \cite{li2012visualizing} as the baseline since it can find variable-length motifs in million-scale time series and utilizes similar hierarchical identification process as the proposed algorithm. We planted 4 motifs of different lengths, 10 instances each, into a random walk time series of length 3 million. The probability that an instance of the motif appears in the time series is smaller than $10^{-5}$, which can be considered rare motif \cite{begum2014rare}). The shape of motifs are generated by using $x=\sum_{i=1}^{5}A_i\sin{\alpha_i x+\beta_i}$
with random parameters $A_i \in [0,5]$, $\alpha_i \in [-2,2]$ and $\beta_i \in [-\pi,\pi]$. The lengths of motifs $L_{M}$ are $1500,3000,6000,12000$. We added 5\% random noise to every instance of the motifs. The mean and variance are also randomly generated. HIME is expected to find at least a pair of non-overlapping subsequences for each motif that highly overlapped with the actual planted instances.

\begin{figure}[h]
 \centering
 \vspace{-2mm}
 \includegraphics[width=70mm]{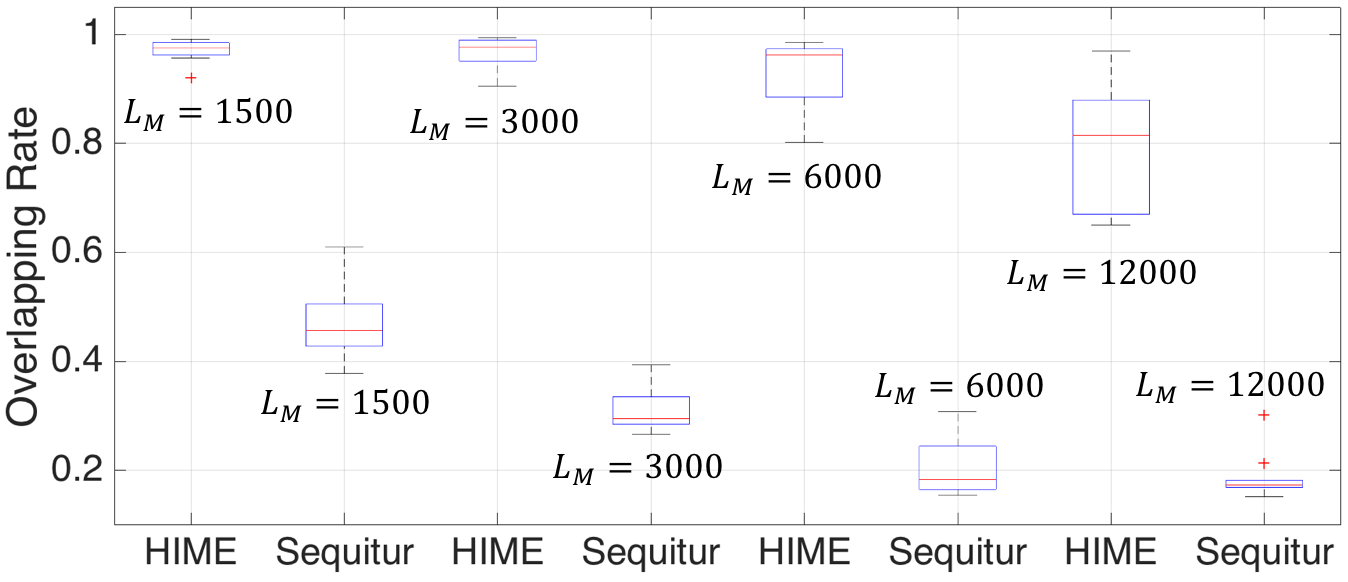}
 \caption{Overlapping rates compared with Sequitur for detecting planted motifs of different lengths}
 \label{fig:exp3}
  \vspace{-6mm}
\end{figure}

The performance of both algorithms is measured by the overlapping rate between the subsequences found and the ground truth locations. Since the grammar-based algorithm also infers motifs based on discretized SAX word sequence, we conduct a grid search for parameter selection of $a=3,4,5,7$ and $w=4,5,7$ (covering the grid search area mentioned in \cite{itr}) and only report the best performance. 

The experimental result is shown in Fig. \ref{fig:exp3}. For Sequitur, the best overlapping rate (0.45 on average) is reached when $L_M=1500$. For all other motif lengths, the overlapping rates are between 0.1 and 0.3. In contrast, HIME consistently gets overlapping rates above 0.8 for motif lengths 1500, 3000 and 6000. The performance decreases when detecting motif of length 12000; however, HIME still can achieve overlapping rates above 0.8 in 6 out of 10 cases for motif length 12000. 

\subsubsection{Performance VS. Number of Instances}

\begin{figure}[h]
 \centering
  \vspace{-2mm}
 %\vspace{-1em}
 \includegraphics[width=70mm]{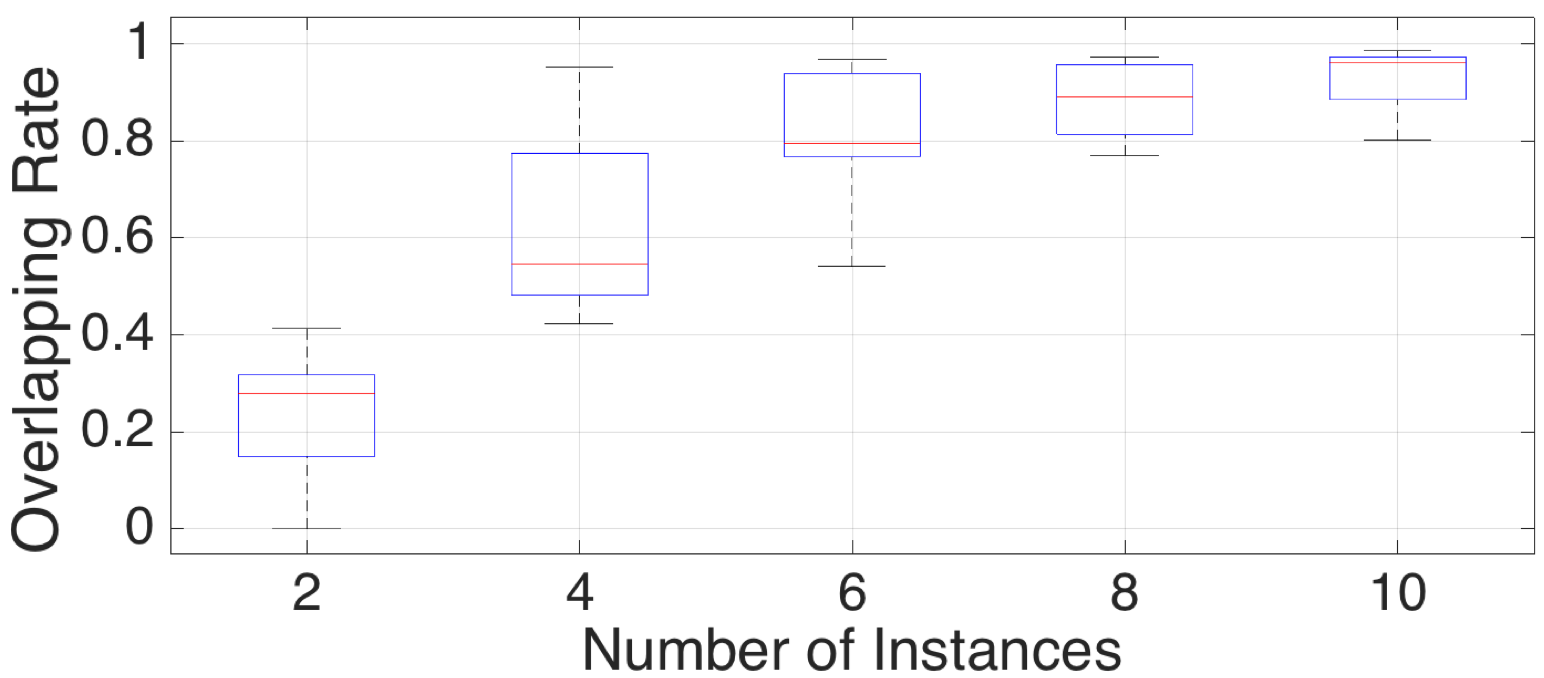}
 \caption{Overlapping rates vs. number of motif instances for motif length 6000}
 \label{fig:l4}
  \vspace{-4mm}
\end{figure}

\begin{table*}[h]
\caption{Scalability Compared with Estimated State-of-the-art Solution (without Numerosity Reduction)}
\centering
\begin{tabular}{ |c|c|c|c|c|c| }
  \hline
   \# of Subsequences & 1 million & 2 million &4 million & 8 million & 16 million\\
   \hline
   Motif Length Range Detected & [300 3020] & [300 3643]& [300 5104] & [300 7775] & [300 10795]\\
   \hline
  STOMP (fixed length) & 2 hr & 16.6 hr & \textcolor{gray}{5.01 days} (est.) & \textcolor{gray}{18.25 days} (est.) & \textcolor{gray}{68 day} (est.)\\
  STOMP $+$ 99\% pruning rate & 2.25 days & 22.7 days & \textcolor{gray}{0.64 yr} (est.) & \textcolor{gray}{3.65 yr} (est.) & \textcolor{gray}{19.9 yr} (est.) \\
  \hline
  HIME & 1.2 min & 2.9 min & 8.4 min & 30 min & 1.6 hr\\
  Post-processing & 0.4 min & 0.8 min & 3.1 min & 15 min & 0.9 hr\\ 
  \hline\hline
  Total & \textbf{1.6 min} & \textbf{3.7 min} & \textbf{11.5 min} & \textbf{45 min} & \textbf{2.5 hr}\\ 
  \hline
\end{tabular}
\label{tab:memory}
\vspace{-2mm}
\end{table*}

\begin{table*}[h]
\caption{Enumeration Range Growth Compared with Sequitur (with Numerosity Reduction)}
\centering
\begin{tabular}{ |c|c|c|c|c|c| }
  \hline
  Time Series Length & 1 million& 2 million& 4 million& 8 million& 16 million\\
  \hline
    Sequitur-Best (Enumeration Range) & 506([300 806]) & 535([300 835]) & 586([300 886]) & 586([300 886]) & 586([300 886])\\
  HIME (Enumeration Range) & \textbf{2307([300 2607])} & \textbf{2731([300 3031])}&\textbf{3705([300 4005])} & \textbf{5965([300 6265])} & \textbf{8811([300 9111])}\\
   \hline
   Sequitur-Best & 3.6 sec. & 12 sec. & 30 sec. & 2 min & 6 min\\
   HIME & 12 sec. & 30 sec. & 1.4 min & 9 min & 40 min\\
  \hline
\end{tabular}
\label{tab:memory}
\vspace{-5mm}
\end{table*}

We conducted an experiment to illustrate the relationship between accuracy and the number of instances repeated on the time series. We use planted motif of length 6000 and adjust the number of instances in each experiment. The overlapping rates vs. the number of instances repeated is shown in Fig. \ref{fig:l4}. According to Fig. \ref{fig:l4}, the accuracy grows rapidly as the number of repeated instances increases. The algorithm can successfully detects the planted motifs with very high overlapping rate (above 80\% overlapped with ground truth) when the number of instances is above 6. Note that a motif only repeating 6 times in million scale time series is considered a rare motif due to the length of time series. Therefore, although HIME is a greedy approximate motif discovery algorithm that does not guarantee exact solution, since the algorithm utilizes all short subsequences overlapped with the instances to detect long motifs, the chance that the motif is detected increases as the number of motif instances increases. Also, motifs often repeat multiple times in a long time series (e.g. if a motif appears with the probability of $10^{-4}$, in a time series of one million in length, it may appear 100 times), so the chance of the motif being detected by HIME is high. 

\subsection{Scalability}

In this subsection, the scalability of algorithm is tested in a 16-million length random walk time series. 

\subsubsection{Execution Time Vs. Data Size}

\begin{figure}[h]    
\centering
    \begin{subfigure}[b]{65mm}
        \includegraphics[width=\textwidth]{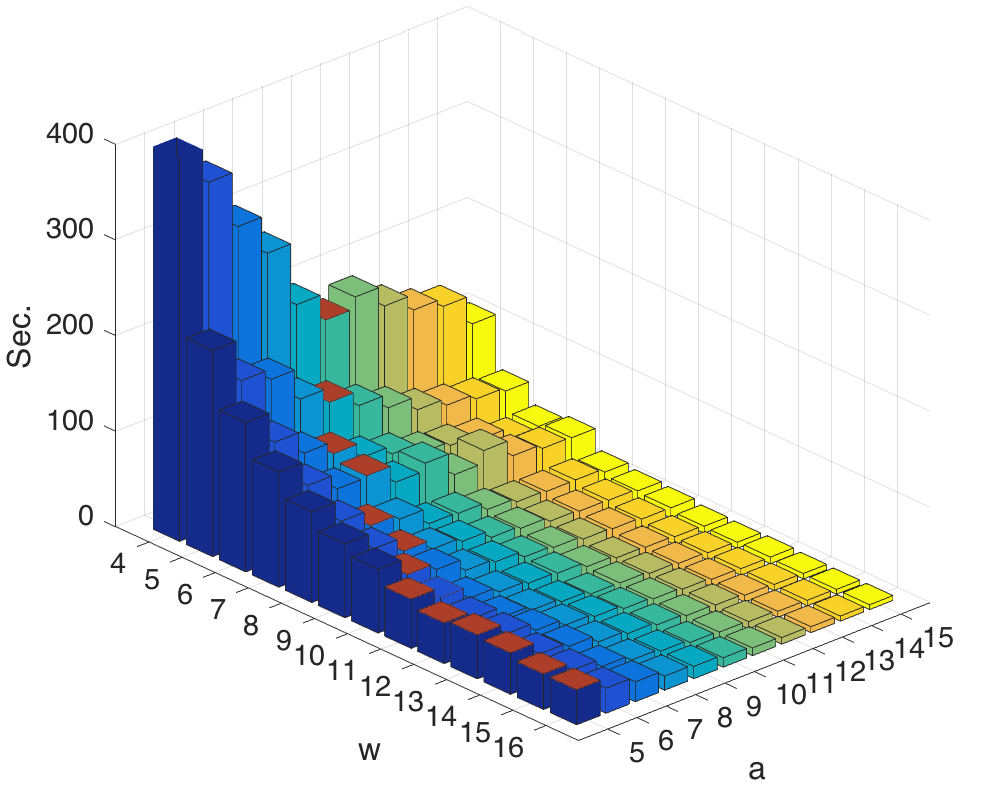}
        \caption{Parameter Vs. Algorithm Execution Time}
        \label{fig:a}
    \end{subfigure}
    \begin{subfigure}[b]{65mm}
        \includegraphics[width=\textwidth]{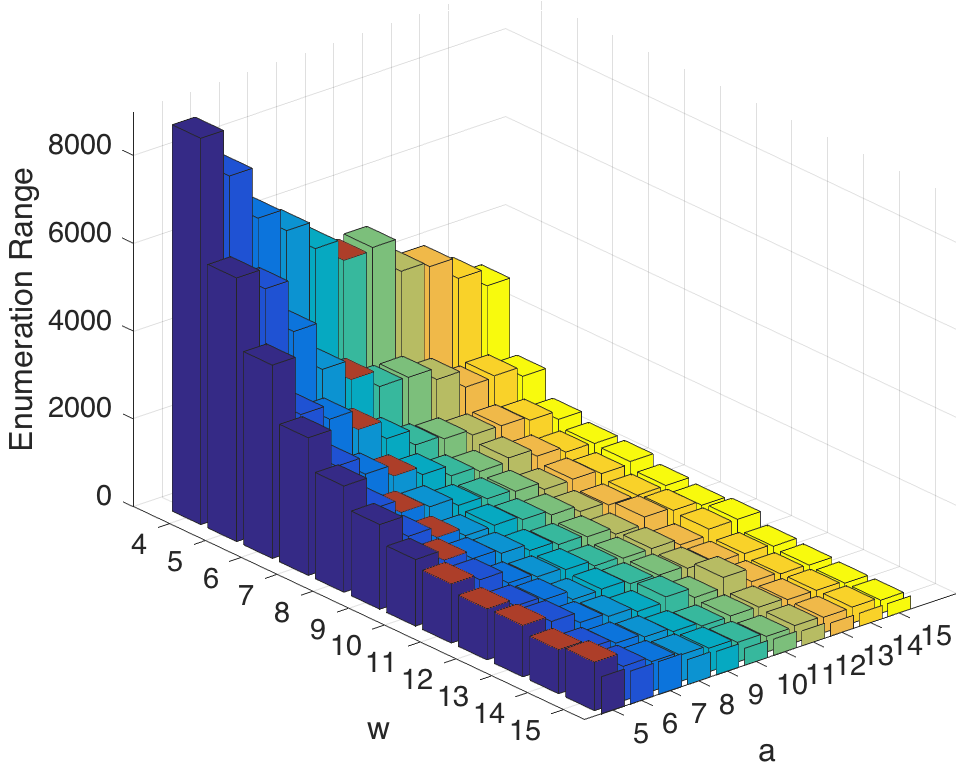}
        \caption{Parameter Vs. Enumeration Range}
        \label{fig:b}
    \end{subfigure}
    \vspace{-2mm}
\caption{Parameter experiments in a random walk time series}
\label{fig:param}
\vspace{-8mm}
\end{figure}

We use STOMP \cite{yeh2016matrix} with 99\% pruning rate to demonstrate the necessity of an approximate approach in million-scale (or larger) time series. In the test case where STOMP takes more than 24 hours, we estimate the execution time by the first 100 iterations (the same estimation approach used in \cite{yeh2016matrix}). The enumeration range of HIME is measured after removing all false positive motifs based on motif threshold function. The result is shown in Table II. The running time for HIME grows significantly slower compared to state-of-the-art algorithms. In the largest test case, HIME takes approximately 2.5 hr to process 16 million subsequences and the length range of motifs found is from 300 to 10795. The estimated execution time for STOMP is 68 days. So even with 99\% pruning rate, it may take 20 years to enumerate the same length range as HIME (though for STOMP, the solution would be exact). Note that GPU version of STOMP \cite{zhumatrix}, which can achieve approximately 150 times speedup, can be used to increase the scalability of STOMP. However, it may still take 47 days. The problem size is simply too large for state-of-the-art algorithms to get the exact solution. In contrast, HIME provides an alternative way to efficiently detect approximate variable-length motifs of large length range in this scale of time series.

\subsubsection{Enumeration Range Vs. Data Size}
We next demonstrate that HIME can efficiently detect motifs in large enumeration range as the length of time series increases. The performance is compared with that of Sequitur (grammar-based approach). Similar to previous experiment, we use grid search to find the best parameters for Sequitur so that the enumeration range can be maximized. In order to make a fair comparison, HIME uses the same numerosity reduction strategy as Sequitur in this experiment instead of the Induction Graph approach introduced in Section IV.D. This way, both algorithms process the same input. The results are shown in Table III. Sequitur's enumeration range stops growing after reaching length 886. In contrast, HIME's enumeration range continues to increase as the time series length grows. Therefore, in large scale time series, HIME can detect significantly longer motifs compared with Sequitur. When the length of time series reaches 16 million, the enumeration range of HIME is one order of magnitude larger than that of Sequitur. We also measure the execution time of both algorithms including post-processing. According to the results, HIME's overall running time is slower than that of Sequitur's. However, considering that HIME's enumeration range is 4-15 times larger than Sequitur's, HIME is more efficient than Sequitur per motif enumeration length. 

 \vspace{-2mm}
\subsection{Parameter Analysis}

In this subsection, we demonstrate that the parameters in HIME are easy to choose. Recall that Algorithm 3 can help determines the alphabet size given a range, so we only need to set one paramter for HIME.

We test HIME with parameter $w$ from 4 to 16 and $a$ from 5 to 15 on a 1 million length random walk time series. The execution time and enumeration range of all parameter combinations are shown in Fig. \ref{fig:param}(a) and Fig. \ref{fig:param}(b) respectively. According to Fig. \ref{fig:param}, HIME's execution time and enumeration range increase as $w$ and $a$ decrease. This is because small $w$ and $a$ allows easier word matching, but loose representation also requires the algorithm to  take extra time to compare and filter out false positives. As the number of distinct SAX words increases with the increase of $w$ and $a$, the execution time and enumeration range of the algorithm is reduced. The parameter combination chosen by Algorithm 3 for different $w$ is labeled in red in the figure. According to the figure, Algorithm 3 tends to select $a$ with trade-off of execution time and enumeration range with equal priority. By using Algorithm 3 to determine $a$, the change in execution time and enumeration range as $w$ increases is the same as the results from setting 2 parameters. So user can select only $w$ unless they want to enumerate a large length range or save time cost due to the length of time series. In these cases, user can set both parameters $w$ and $a$ to balance the enumeration range and execution time. %based on their expectation.

\vspace{-1mm}
\subsection{Case Studies}
%\vspace{-1mm}

In this section, we show that HIME can find high quality motifs in several real world million scale time series data. As demonstrated in previous section, existing algorithms are too costly to be applied on problems of such scale. 

\subsubsection{Variable Length Similar Subsequences Discovery in DNA Sequence}

As shown in previous work, converting DNA sequence to time series \cite{rakthanmanon2012searching} can help researchers understand structural similarity \cite{zhumatrix}\cite{rakthanmanon2012searching} between DNA subsequences. In this case study, we use HIME to find repeating subsequences in Human hgY chromosome. We converts hgY to a 26 million length time series based on the algorithm described in \cite{rakthanmanon2012searching}. We set PAA size $w$ to $10$ so that the algorithm can finish the search process within half an hour. We set minimum motif length $l$ to 1000. 

\begin{figure}[h]
 \centering
 \vspace{-2mm}
 \includegraphics[width=80mm]{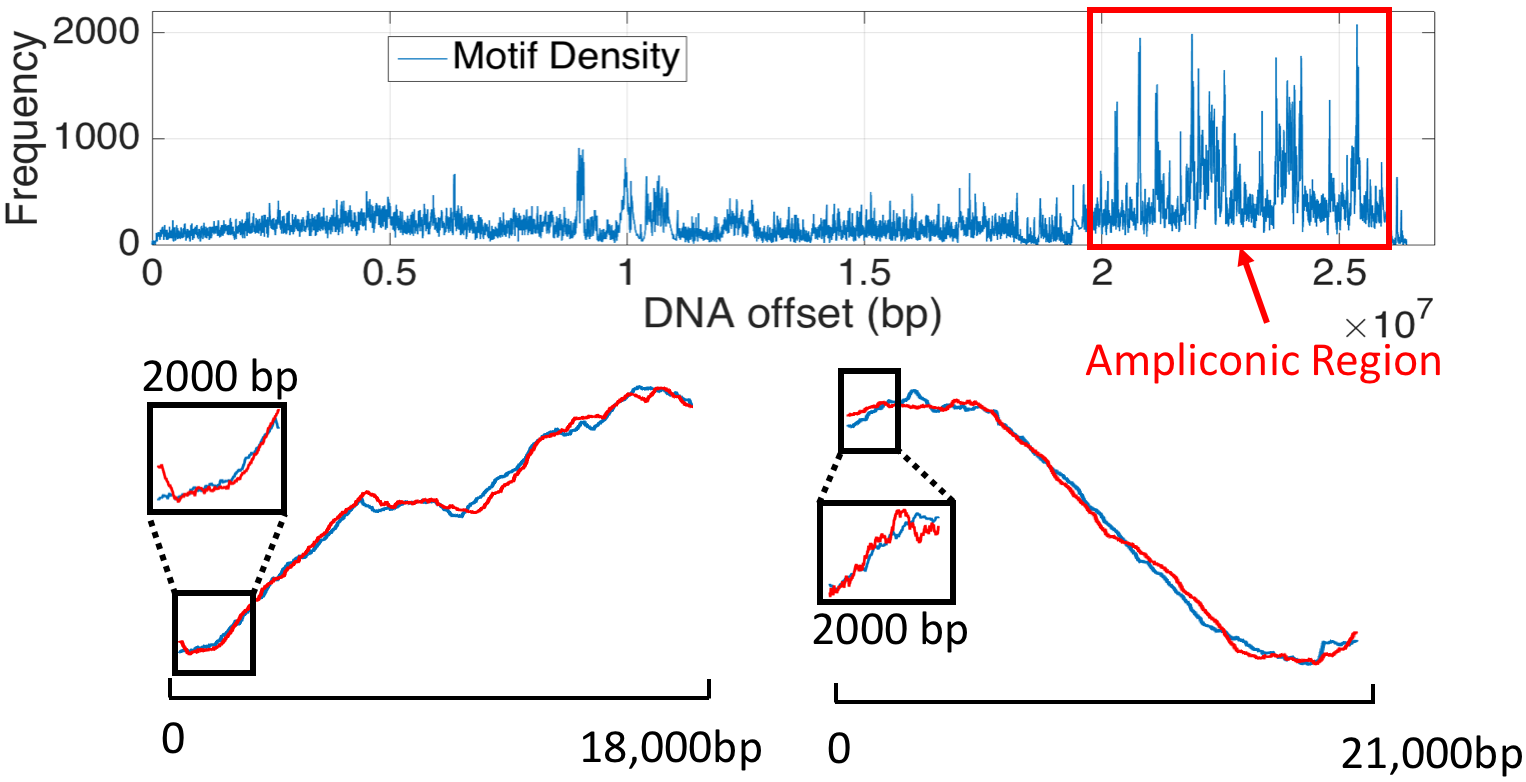}
 \caption{Examples of time series motifs found in human Y-Chromosome}
 \label{fig:dna}
  \vspace{-7mm}
\end{figure}

The motif density curve \cite{senin2014grammarviz} is shown in Fig. \ref{fig:dna}.top. We observe a region that has a large amount of repeated subseqeunces. According to \cite{skaletsky2003male}, this region is the longest ``ampliconic region'' (labeled in red) in Y chromosome. As explained in \cite{skaletsky2003male}, ampliconic region, or \textit{``the ampliconic segments, are composed largely of sequences that exhibit marked similarity —as much as 99.9\% identity over tens or hundreds of kilobases—to other sequences in the MSY''}

Two examples of long motifs discovered by our algorithm are shown in Fig. \ref{fig:dna}.bottom. The two motifs found indicate similarity between 2 long DNA subsequences of length 18,000 and 22,000, respectively. Note that previous work \cite{skaletsky2003male} set a sliding window of length 2000 to find similar DNA sequence patterns.  We show the first 2000 points of the long motifs in the black box. The respective intra-distances between both pairs of subseqeunces are greater than the motif threshold. In other words, they would not be discovered by fixed-length motif discovery algorithm with motif length 2000. The experiment indicates that, by enabling variable-length motif discovery in long time series, HIME provides an opportunity to discover potentially surprising patterns that are hard to be detected.

\subsubsection{Discovering motifs in Bird Soundtrack}

Existing work \cite{begum2014rare}\cite{shokoohi2015discovery} shows that motifs can be used to find repeated calls from bird soundtrack. In this experiment, we test our approach with 600 records of Rufours Capped, Spix and Azard spinetail birds in \cite{xeno}. We use the second Mel-Frequency Coefficient (MFCC) with 250 Hz to form a 5 million length time series. HIME is applied to this time series with minimum length equal to 0.5 sec. 

\begin{figure}[h]    
\vspace{-3mm}
\centering
    \begin{subfigure}[b]{75mm}
        \includegraphics[width=\textwidth]{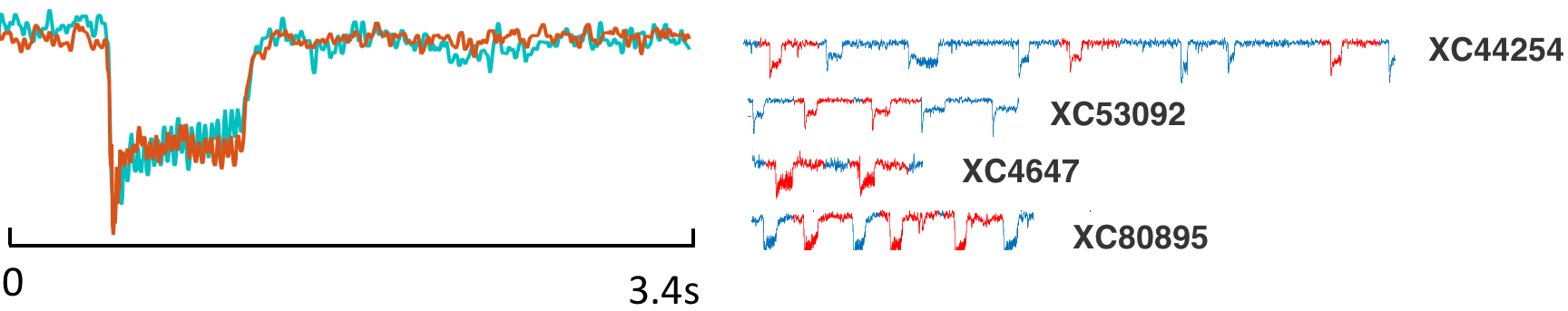}
        \caption{Rufous Capped Spinetail}
        \label{fig:a}
    \end{subfigure}
    \begin{subfigure}[b]{75mm}
        \includegraphics[width=\textwidth]{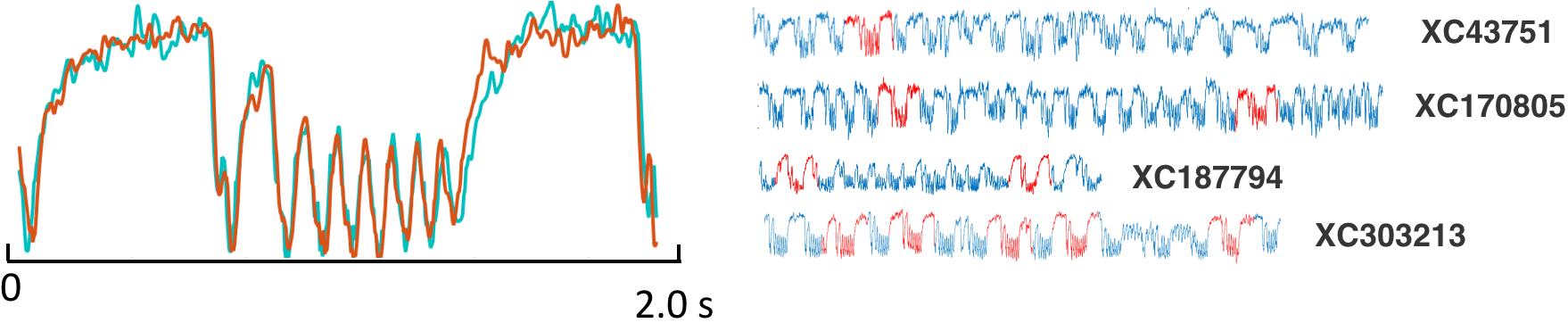}
        \caption{Spix Spinetail}
        \label{fig:b}
    \end{subfigure}
    \begin{subfigure}[c]{75mm}
        \includegraphics[width=\textwidth]{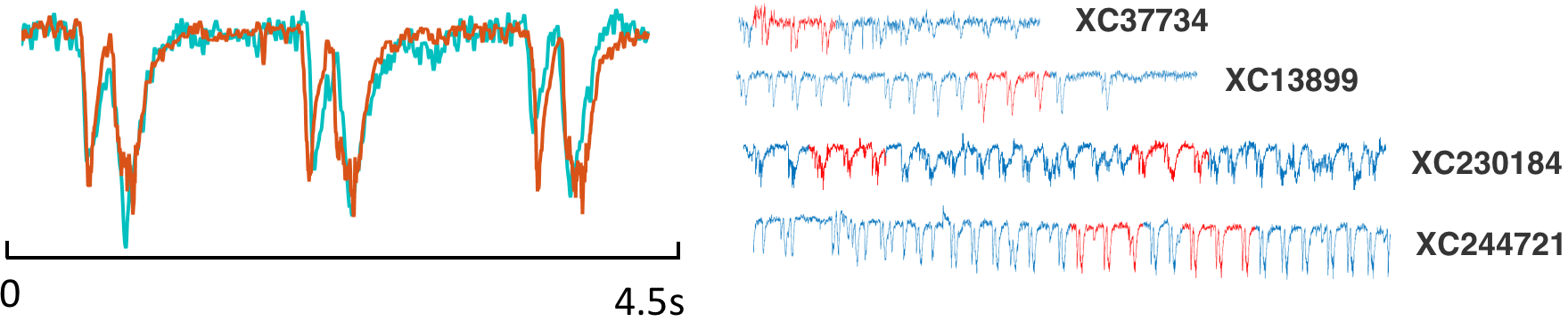}
        \caption{Azard Spinetail}
        \label{fig:c}
    \end{subfigure}
    \vspace{-2mm}
\caption{Examples of motifs detected by HIME in Bird Soundtrack}
\label{fig:bird}
\vspace{-5mm}
\end{figure}

Three examples of motifs discovered with 3.4, 2.0 and 4.5 seconds are shown in Fig. \ref{fig:bird}. By using the query algorithm \cite{FastestSimilaritySearch} to retrieve all motif instances from the closest pair detected, we find that the similar subsequences of these three motifs refer to the same species of birds. Four records of each bird species are shown in Fig. \ref{fig:bird}.right. The locations of the motifs are labeled in red. These 3 motifs also explain some unique sound patterns for each bird. For example, Fig. \ref{fig:bird}(a) shows that Rufous Capped may have a silent 1.5 sec after a bird call. Fig. \ref{fig:bird}(b) shows that between every call of Spix Spinetail, it tends to have a 0.5 sec. gap. Fig. \ref{fig:bird}(c) indicates that Azard Spinetail often generates three consecutive bird calls. Such information provides useful insights that can help researchers understand bird behaviors.

\subsubsection{Revealing Rare Patterns in Long Electric Power Usage Time Series}

\begin{figure}[h]
 \centering
 \vspace{-3mm}
 \includegraphics[width=70mm]{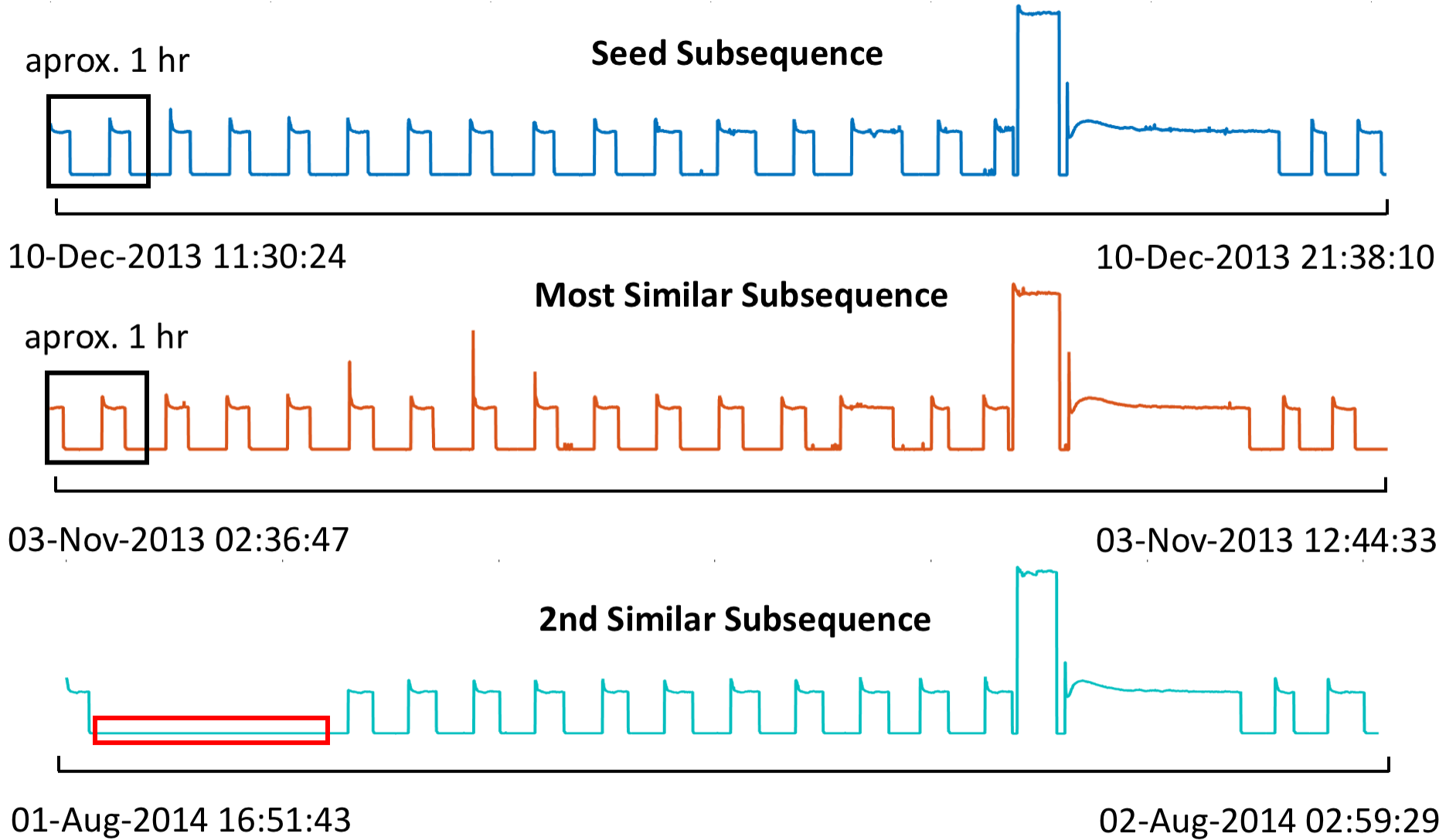}
 \caption{Example of rare motif in electric power usage data}
 \label{fig:3m}
\end{figure}

In this case study, we show that by enabling large range of motif discovery, HIME can detect long and rare patterns in real world time series. HIME is applied on a 7.4 million length freezer electric power usage time series recorded in \cite{murray2015energy} with minimum length approximately equal to 1 hour period. An example of long pattern is shown in Fig. \ref{fig:3m}. HIME detects the repeating subsequences shown in Fig. \ref{fig:3m} (top, middle; approximately 10 hours). Using the query algorithm \cite{FastestSimilaritySearch}, we find that even the $2^{nd}$ similar subsequence (Fig.\ref{fig:3m} (bottom)) already has obvious sequential difference (difference is surrounded in red box). This may indicate that this is a rare motif. Besides, two subsequences found by HIME are 1 month and 7 days apart in the time series, hence hard to find in shorter time series.

 \vspace{-7mm}
\section{Conclusion}

We introduce a new algorithm named HIME, to detect time series motifs of large range of different lengths in million-scale time series that state-of-the-art algorithms cannot handle efficiently. HIME can process million subsequences with execution time less than 1 minute, which is much faster than any existing algorithm to date. Compared with sequence matching based approach such as grammar induction-based motif discovery algorithms, HIME achieves one order of magnitude increase in search range. In the case studies, we demonstrate that the motifs found by HIME are meaningful and can potentially have significant impacts in various applications.
 \vspace{-1mm}
\bibliographystyle{abbrv}

\end{document}